\documentclass[conference]{IEEEtran}
	\usepackage{amsmath,amsfonts}
	\usepackage{bm}
	\usepackage{algorithmic}
	\usepackage{hyperref}
	\usepackage{algorithm}
	\usepackage[caption=false,font=normalsize,labelfont=sf,textfont=sf]{subfig}
	\usepackage{textcomp}
	\usepackage{stfloats}
	\usepackage{url}
	\usepackage{verbatim}
	\usepackage{epsfig}
	\usepackage{cite}
	\usepackage{color}
	\usepackage{multirow,makecell}
	\usepackage{amssymb}
	\usepackage{graphicx}
\usepackage{algorithm}
	\def\BibTeX{{\rm B\kern-.05em{\sc i\kern-.025em b}\kern-.08em
			T\kern-.1667em\lower.7ex\hbox{E}\kern-.125emX}}
\begin{document}

\title{Primary Rate Maximization in Movable Antennas Empowered Symbiotic Radio Communications}

\author{\IEEEauthorblockN{Bin Lyu\IEEEauthorrefmark{1},  Hao Liu\IEEEauthorrefmark{1}, Wenqing Hong\IEEEauthorrefmark{1}, Shimin Gong\IEEEauthorrefmark{2}, and Feng Tian\IEEEauthorrefmark{1}}
\IEEEauthorblockA{\IEEEauthorrefmark{1}
Nanjing University of Posts and Telecommunications, Nanjing 210003, China}
\IEEEauthorblockA{\IEEEauthorrefmark{2}  Sun Yat-sen University, Shenzhen 518055,  China}
}
		
\markboth{IEEE Communications Letters}%
{Shell \MakeLowercase{\textit{et al.}}: A Sample Article Using IEEEtran.cls for IEEE Journals}

\maketitle
		
\begin{abstract}
In this paper, we propose a movable antenna (MA) empowered scheme for symbiotic radio (SR) communication systems. Specifically, multiple antennas at the primary transmitter (PT) can  be flexibly  moved to favorable locations to boost the channel conditions of the primary and secondary transmissions. The primary transmission is achieved by the active transmission from the PT to the primary user (PU), while the backscatter device (BD) takes a ride over the incident signal from the PT to passively send the secondary signal to the PU. Under this setup, we consider a primary  rate maximization problem by jointly optimizing the transmit beamforming and the positions of MAs at the PT under a practical bit error rate constraint on the secondary transmission. Then, an alternating optimization framework with the utilization of the successive convex approximation, semi-definite processing and simulated annealing (SA) modified particle swarm optimization (SA-PSO) methods is proposed to find the solution of the transmit beamforming and MAs' positions. Finally, numerical results are provided to demonstrate the performance improvement provided by the proposed MA empowered scheme and the proposed algorithm.
			
\end{abstract}
		

\section{Introduction}
		
\IEEEPARstart{W}{ith} the emergence of 6G communications, the requirements on spectrum and energy efficiency  are becoming more and more stringent \cite{LiangSurvey}. On one hand, 6G communications will  facilitate the ubiquitous deployment of wireless devices, massive access to which poses a great challenge to the scarce spectrum resources. On the other hand, the radio-frequency (RF) components used in wireless devices are energy-hungry, which  is most unlikely to attain satisfactory energy efficiency performance. The two drawbacks may seriously restrict the application of 6G communications in the near future. Thus, how to overcome the drawbacks is worthy of investigations. 

Recently, symbiotic radio (SR) communications \cite{LongIoT} has been emerged as a promising solution to the above drawbacks. In a typical SR  system, wireless devices from the secondary system can reuse the spectrum band allocated to the primary transmission and take a ride over the primary signals for passive information delivery. Thus, these  devices do not need an additional spectrum band and are unnecessary to equip with RF components. This setup boosts the system spectrum and energy efficiency and is friendly to 6G communications. Motivated by this advantage, SR has been widely investigated \cite{LongIoT,QQZhang,ZChu,Zeng}. In \cite{LongIoT}, the parasitic SR (PSR) paradigm was designed to adapt the diversified communication requirements. 
In \cite{QQZhang}, the duration ratio between the primary and secondary signals to attain a mutualistic relationship for SR communications was investigated.  In \cite{ZChu}, the impact of finite blocklength channel codes on the secondary transmission was studied. In \cite{Zeng}, the cell-free paradigm was proposed for SR communications to avoid the interference caused by the cellular network structure and extend network coverage. However, the system performance of \cite{QQZhang,ZChu,Zeng}, especially the secondary transmission, was not satisfactory.  One important reason for this phenomenon is that  the antennas at the primary transmitter (PT) in \cite{QQZhang,ZChu,Zeng} was located on fixed positions, based on which the degrees of freedom (DoFs) for achieving high spatial diversity cannot be fully leveraged. 

To address the challenge above, an emerging concept, movable antenna (MA), has been proposed as an efficient approach since it can make full use of the spatial DoFs \cite{MAMag,MATWC,PSOMA}. Unlike  conventional antennas with fixed positions, the flexible cable is utilized to establish the connection between each MA and its corresponding RF chain, enabling the position of each MA to be dynamically  adjusted to construct a favorable wireless environment \cite{MAMag}. Currently, the investigation of MA empowered wireless communications is in its quite early stage.  In  \cite{MATWC}, the positions of MAs at the transmitter and receiver were jointly optimized to capture the ultimate downlink system capacity. The authors in \cite{PSOMA} shifted their attention to  MA based uplink communication systems with multiple users.  The results in \cite{MATWC} and \cite{PSOMA} demonstrated that the application of MAs can achieve amazing system performance improvement. 

However, how to apply MAs to handle the unsatisfactory communication performance in SR systems has not been solved, especially when both primary and secondary transmissions are taken into account. To fill in this research gap, in this paper, we propose an MA empowered scheme for the SR communication system, which comprises of a PT, a backscatter device (BD) and a primary user (PU). The PT is equipped with MAs to send primary signals to the PU with a fixed antenna (FA). By considering the PSR paradigm, the FA-based BD takes a ride over the primary signals for delivering its own information to establish the secondary transmission. Under this setup,
the MAs at the PT are utilized to wholly explore the DoFs for boosting the primary and secondary transmissions. 
Moreover, we investigate a primary transmission rate maximization problem under the bit error rate (BER) constraint on the secondary transmission. To handle the non-convexity faced by the original problem, we propose an alternating optimization (AO) framework to decompose it into two sub-problems. For the sub-problem of designing the transmit beamforming, the successive convex approximation (SCA) and semi-definite processing (SDP) are both leveraged to find a near-optimal solution. While, the simulated annealing (SA) modified particle swarm optimization (SA-PSO) method is utilized to solve the sub-problem of optimizing the positions of MAs. Finally, numerical results are conducted to demonstrate the superior performance provided by utilizing the MAs at the PT and the proposed algorithm with the SA-PSO.

\section{System Model}
\begin{figure}
  \centering
  \includegraphics[width=0.43\textwidth]{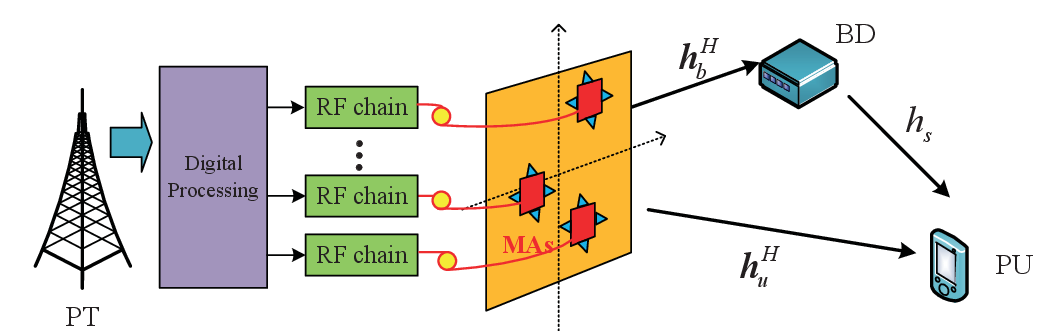}
  \caption{The MA empowered SR communication system.}
  \label{fig:system_model}
\end{figure}

We consider an MA empowered SR communication system, in which there exist a PT with $K$ MAs, a BD equipped  with single FA, and a PU equipped  with single FA. The illustration of the system model is shown in Fig. \ref{fig:system_model}.
 For the $k$-th MA at the PT, it utilizes a flexible cable to connect with its corresponding RF chain for ensuring the movement. The position of the $k$-th MA is denoted by $\bm{p}_k =[x_k, y_k]^T$, which can be updated in a two-dimensional region with size  $[x_\text{min},x_\text{max}] \times [y_\text{min}, y_\text{max}]$. While the positions of the FAs at the BD and the PU are static and represented by 
$\bm{p}_b = [x_b,y_b]^T$ and $\bm{p}_u = [x_u,y_u]^T$, respectively.
Under this setup, the channel condition between the PT and the BD and that between the PT and the PU can be improved by moving the MAs to  favorable positions. 

\subsection{Channel Model}
We consider a quasi-static fading channel model, based on which the channels remain static in each block when the MAs  are fixed. We first describe the channels related with the PT. Let $L^t_p$  represent the amount of transmit paths between the PT and the node-$\kappa$, where $\kappa =\{b,u\}$,
$\kappa=b$ represents that the node is the BD, and $\kappa=u$ represents that the node is  the PU. 
 The azimuth and elevation angles of departure (AoDs) of the $m$-th transmit path between the PT and the node-$\kappa$ are presented by $\phi_{m}^t$ and $\theta_{m}^t$, respectively, where $m \in \mathcal{L}^t $, and $\mathcal{L}^t = \{1,\ldots,L^t_p\}$ is the set of the PT's transmit paths. 
 According to \cite{MATWC}, the signal propagation difference between the $m$-th transmit path of the $k$-th MA and the origin of the transmit region, denoted by $\rho_{m}^t(\bm{p}_k)$, is expressed as 
\begin{align}
\rho_{m}^t(\bm{p}_k) &= x_k \sin\theta_{m}^t \cos\phi_{m}^t + y_k \cos \theta_{m}^t,~ 
\end{align}
where $k \in \mathcal{K},~m \in \mathcal{L}^t$, and $\mathcal{K}= \{1,\ldots,K\}$ is the set of the MAs at the PT.

 Similarly, for $n \in \mathcal{L}_\kappa^r$, the $n$-th receive path's azimuth and elevation angles of arrival (AoAs) between the PT and the node-$\kappa$ are  denoted by  $\phi_{\kappa,n}^r$ and $\theta_{\kappa,n}^r$, respectively, where  $\mathcal{L}_\kappa^r= \{1,\ldots, L_\kappa^r\}$ is the set of the receive paths associated with the node-$\kappa$, and $L_\kappa^r$ is the amount of corresponding receive paths. Denote the signal propagation difference between the $n$-th receive path of the node-$\kappa$  and the origin of the receive region as $\rho^r_{\kappa,n}(\bm{p}_\kappa)$, which is formulated as
\begin{align}
 \rho_{\kappa,n}^r(\bm{p}_\kappa) &= x_\kappa \sin\theta_{\kappa,n}^r \cos\phi_{\kappa,n}^r + y_\kappa \cos \theta_{\kappa,n}^r,
\end{align}
where $\kappa \in \{b,u\},~n \in \mathcal{L}_\kappa^r$.

The channel vector from the PT to the node-$\kappa$, denoted by $\bm{h}_{\kappa}^H$, is given by
\begin{align}
\bm{h}_{\kappa}^H = \bm{f}_\kappa^H \bm{\Sigma}_\kappa \bm{G},~\kappa \in \{b,u\},
\end{align}
where $\bm{f}_\kappa = [e^{j \frac{2\pi}{\lambda}\rho_{\kappa,1}^r}, \ldots, e^{j \frac{2\pi}{\lambda}\rho_{\kappa,L_\kappa^r}^r}  ]^T \in \mathbb{C}^{L_\kappa^r \times 1} $ is the receive field response vector at the node-$\kappa$,  $\bm{\Sigma}_\kappa \in \mathbb{C}^{L_\kappa^r \times L_p^t}$ is the path-response matrix, $\bm{G}= [ \bm{g}^t(\bm{p}_1), \ldots, \bm{g}^t(\bm{p}_K) ] \in \mathbb{C}^{ L_p^t \times K }$ is the transmit field response matrix, and $\bm{g}^t (\bm{p}_k) = [e^{j \frac{2\pi}{\lambda}\rho_{1}^t}, \ldots, e^{j \frac{2\pi}{\lambda}\rho_{L_p^t}^t}  ]^T \in \mathbb{C}^{L_p^t \times 1}$ is the transmit field response vector  associated with the $k$-th MA, and $\lambda$ is the  carrier wavelength. It is noted that $\bm{f}_\kappa$ is constant due to the static position of the node-$\kappa$, and $\bm{G}_\kappa $ is a variable due to the movable characteristic of the antennas at the PT.

Then, we model the channel between the BD and the PU, which is represented by $h_{s}$. Similarly, $h_{s}$ can be expressed as 
\begin{align}
h_{s} =  \bm{f}_{u}^H \bm{\Sigma}_s  \bm{g}_s,
\end{align}
where   $\bm{\Sigma}_s \in \mathbb{C}^{L_s^r \times L_s^t}$ is the path-response matrix, $\bm{g}_{s} = [e^{j \frac{2\pi}{\lambda}\rho_{s,1}^t}, \ldots, e^{j \frac{2\pi}{\lambda}\rho_{s,L_s^t}^t}  ]^T \in \mathbb{C}^{ L_s^t\times 1}$ is the transmit field response at the BD,  $L_s$ is the amount of transmit paths between the BD and the PU, $\rho_{s,\varrho}^t = x_b \sin \theta_{s,\varrho}^t \cos\phi_{s,\varrho}^t + y_b \cos \theta_{s,\varrho}^t$, $\theta_{s,\varrho}^t$  represents the  elevation AoD, $\phi_{s,\varrho}^t$ represents the azimuth AoD,  and $\varrho =1,\ldots, L_s^t$. Since both the BD and PU are equipped with the FA, $h_{s}$ is also a constant.

\subsection{Transmission Model}
In the considered system, the PT sends primary signals to the PU with the involvement of the BD.  Denote the primary signal at the PT as $\bm{w}s(l)$, where $s(l)$ is the primary signal with unit power,  $\bm{w}\in \mathbb{C}^{K \times 1}$ is the transmit beamforming vector satisfying $||\bm{w}||^2 \le P_\text{max}$, and $P_\text{max}$ is the maximum power provided by the PT. The incident signal at the BD from the PT is denoted by $\bm{h}^H_b \bm{w}s(l)$. By catching a ride over this incident signal, the BD can send its own information-carrying signal $c(l)$, which is also named as the secondary signal. Considering the implementation of the BD in reality, we adopt the binary phase shift keying modulation, i.e., $c(l) \in \{-1,1\}$.
The received signal at the PU is formulated as \cite{LongIoT} 
\begin{align}
\label{ReceivedSignal}
y(l) = \bm{h}_u^H \bm{w}s(l) + \sqrt{\alpha} {h}_s \bm{h}_b^H \bm{w} s(l) c(l)+z(l),
\end{align}
where $\alpha \in [0,1]$ is the reflection efficiency at the BD, $z(l) \sim \mathcal{CN}(0,\sigma^2)$ is the additive Gaussian white noise at the PU. The first term of \eqref{ReceivedSignal} is the signal transmitted through the direct link, and the second term represents the reflected signal from the BD combing $s(l)$ and $c(l)$.

As the PSR paradigm is considered, the PU first decodes $s(l)$ by treating the second term as the interference \cite{LongIoT}. Accordingly, the expression of the signal to interference plus noise ratio (SINR)  is 
$\gamma_p = \frac{|\bm{h}_u^H \bm{w}|^2 }{\alpha |h_s|^2 |\bm{h}_b^H \bm{w}|^2 + \sigma^2 }.$
Then, the primary  rate  from the PT to the PU is 
\begin{align}
R_p = \log_2(1+\gamma_p).
\end{align}

After decoding $s(l)$ and removing the first term from \eqref{ReceivedSignal}, the signal to noise ratio (SNR) of decoding $c(l)$ is given by
$\gamma_c = \frac{\alpha |h_s|^2 |\bm{h}_b^H \bm{w}|^2 }{\sigma^2}$. According to \cite{Hua} and \cite{DCBook}, the average BER for decoding $c(l)$ is then expressed as $e_c = \frac{1}{2} -\frac{1}{2} \sqrt{\frac{\gamma_c}{1+\gamma_c}}.$

\section{Primary  rate maximization}
In this section, we investigate the primary  rate maximization problem under the BER constraint for decoding the secondary signal. To be specific, the maximization problem is formulated as a joint optimization of the transmit beamforming and the positions of the MAs at the PT, which is formulated as 
\begin{equation}\tag{$\textbf{P1}$} 
		\begin{aligned}
			\max_{\bm{w}, \bm{p}} ~~&  R_p \\ 
			\text{s.t.}~~ & \text{C1:~} ||\bm{w} ||_2^2 \le P_\text{max}, \\
			& \text{C2:~} e_c \le  e_\text{max}, \\
			& \text{C3:~} x_\text{min}\le x_k \le x_\text{max},\\
			  &~~~~~y_\text{min}\le y_k \le y_\text{max},~ k \in \mathcal{K},\\
			& \text{C4:~} \left\|\bm{p}_k - \bm{p}_\iota \right \|_2 \ge d_\text{min},~k, \iota \in \mathcal{K},~k\neq \iota, \\
		\end{aligned}
	\end{equation}
	where $\bm{p}= [\bm{p}_1^T,\ldots, \bm{p}_K^T]^T$,
	 $e_\text{max}$ is the maximum allowable BER for decoding $c(l)$, $d_\text{min}$ represents the minimum distance between any two MAs. C1 indicates that the transmit power at the PT is constrained by its available maximum power $P_\text{max}$, C2 is the BER constraint at the PU for ensuring the quality of service of the secondary transmission, C3 restrains the movable region of the MAs, and C4 is the distance constraint to avoid the coupling effect on MAs.

	It is observed that \textbf{P1} is a highly non-convex problem since the objective function and the constraints C1, C2 and C4 are all non-convex, especially the coupled variables in the objective function and  C2. To make this challenging problem tractable, an alternating optimization (AO) framework is proposed, in which $\bm{w}$ and $\bm{p}$ are updated in an alternating manner. Specifically, the  SCA and SDP methods are explored for the optimization of $\bm{w}$ to guarantee an accurate solution, and the SA-PSO method proposed in \cite{HPSO} is applied to optimize the positions of MAs.

	\subsection{Optimization of transmit beamforming}
	\label{OpTB}
We first fix the positions of MAs and design $\bm{w}$ to maximize $R_p$ by solving the following sub-problem
	\begin{equation}\tag{$\textbf{P2}$} 
		\begin{aligned}
			\max_{\bm{w}} ~~&  R_p \\ 
			\text{s.t.}~~ & \text{C1:~} ||\bm{w} ||_2^2 \le P_\text{max}, \\
			& \text{C2:~} e_c \le  e_\text{max}.
		\end{aligned}
	\end{equation}
	Before solving \textbf{P2}, we first analyze the structure of the constraint C2. It is easy to prove that $e_c$  is a monotonically decreasing function associated with $\gamma_c$. Thus, C2 can be equivalently transformed as 
$\bar{\text{C2}}:~\gamma_c \ge \gamma_\text{min}$, where $\gamma_\text{min}$ is the minimum SNR for ensuring the  BER constraint and can be obtained under the condition $e_c = e_\text{max}$ by using the bisection method. With $\bar{\text{C2}}$, we then rewrite \textbf{P2} as 
		\begin{equation}\tag{$\textbf{P2.1}$} 
		\begin{aligned}
			\max_{\bm{w}} ~~&  R_p \\ 
			\text{s.t.}~~ & \text{C1},~\bar{\text{C2}}.
		\end{aligned}
	\end{equation}
	To handle the non-convexity of \textbf{P2.1}, we first introduce $\bm{W}=\bm{w}\bm{w}^H$ for reformulation. It is obvious $\bm{W}$ is a rank-one matrix, i.e., $\text{Rank} (\bm{W}) = 1$. By utilizing $\bm{W}$, the constraints C1 and $\bar{\text{C2}}$ are reformulated as 
	\begin{align*}
&\tilde{\text{C1}}:~ \text{Tr} (\bm{W}) \le P_\text{max}, \\
&\tilde{\text{C2}}:~\frac{\alpha |h_s|^2 \text{Tr} (\bm{h}_b \bm{h}_b^H \bm{W})  }{\sigma^2} \ge \gamma_\text{min}, 
	\end{align*}
which are both affine.
	Similarly, $R_p$ can be  transformed as 
	\begin{align}
	\label{UpdatedRate}
R_p = \log_2 \left(1+ \frac{ \text{Tr}(\bm{h}_u \bm{h}_u^H \bm{W})  }{\alpha |h_s|^2 \text{Tr} (\bm{h}_b \bm{h}_b^H \bm{W}) +\sigma^2 } \right).
	\end{align}
	However, $R_p$ defined in \eqref{UpdatedRate} is still non-convex due to  the fractional form. To deal with this difficulty, the SCA method is utilized to derive the lower-bound of $R_p$, which is given by 
	\begin{align}
R_p &\ge \log_2(\text{Tr}(\bm{h}_u \bm{h}_u^H \bm{W}) + \alpha |h_s|^2 \text{Tr} (\bm{h}_b      \bm{h}_b^H \bm{W}) +\sigma^2 ) \nonumber \\
&  - \log_2 (\alpha |h_s|^2 \text{Tr} (\bm{h}_b      \bm{h}_b^H \bm{W}^{(\varsigma)}) +\sigma^2) \nonumber \\
& - \frac{\alpha |h_s|^2 \text{Tr} [\bm{h}_b      \bm{h}_b^H (\bm{W} -\bm{W}^{(\varsigma)})] }{\ln(2) [\alpha |h_s|^2 \text{Tr} (\bm{h}_b      \bm{h}_b^H \bm{W}^{(\varsigma)}) +\sigma^2]  } \triangleq \tilde{R}_p,
	\end{align}
	where $\bm{W}^{(\varsigma)}$ is a feasible point of $\bm{W}$ in the $\varsigma$-th iteration of implementing the SCA method.
	It can be found that $ \tilde{R}_p$ is a  concave function associated with  $\bm{W}$. With the updated objective function and constraints and relaxing the constraint $\text{Rank}(\bm{W}) = 1$, $\textbf{P2.1}$ can be transformed into the following SDP problem 
		\begin{equation}\tag{$\textbf{P2.2}$} 
		\begin{aligned}
			\max_{\bm{W}} ~~&  \tilde{R}_p \\ 
			\text{s.t.}~~ & \tilde{\text{C1}},~\tilde{\text{C2}}.
		\end{aligned}
	\end{equation}
The optimal solution  to \textbf{P2.2},  denoted by $\bar{\bm{W}}$, can be attained by utilizing the interior-point method. According to \cite{TWC}, it is proved that $\bar{\bm{W}}$ is a rank-one matrix, which ensures the optimality  of the obtained solution from \textbf{P2.2} after removing the rank-one constraint. 

By iteratively solving \textbf{P2.2} until the convergence, the optimal transmit beamforming matrix is finally obtained and denoted by $\bm{W}^*$. Then, the optimal solution of \textbf{P2.1}, denoted by $\bm{w}^*$, is attained by implementing the singular value decomposition of $\bm{W}^*$.

\subsection{Optimization of  the MAs' positions}
\label{OpMA}
With the attained solution in Section \ref{OpTB}, we then find the optimal positions of the MAs by solving \textbf{P3} as
\begin{equation}\tag{$\textbf{P3}$} 
		\begin{aligned}
			\max_{\bm{p}} ~~&  R_p \\ 
			\text{s.t.}~~ & \text{C1},~\bar{\text{C2}},~\text{C3},~\text{C4}.
		\end{aligned}
	\end{equation}
	It is generally challenging to solve \textbf{P3}  due to the huge solution space, i.e., $[x_\text{min},x_\text{max}]^K \times [y_\text{min}, y_\text{max}]^K$, which results in an extremely high computational complexity. To handle this difficulty and guarantee the solution accuracy,  the SA-PSO based method \cite{HPSO} is applied as an effective approach. By using the SA-PSO method, we first introduce $S$ particles, whose initial positions and velocities are represented by $\mathcal{P}^{(0)} = \{\bm{p}_1^{(0)}, \bm{p}_2^{(0)},\ldots, \bm{p}_S^{(0)} \}$ and $\mathcal{V}^{(0)} = \{\bm{v}_1^{(0)}, \bm{v}_2^{(0)},\ldots, \bm{v}_S^{(0)} \}$, respectively, and $\bm{p}_s^{(0)} = [\bm{p}_{s,1}^{(0)}, \bm{p}_{s,2}^{(0)},\ldots, \bm{p}_{s,K}^{(0)}]$. $\bm{p}_s^{(0)}$ denotes a possible solution of the  positions with $\bm{p}_{s,k}^{(0)}= [ x_{s,k}^{(0)}, y_{s,k}^{(0)} ]$, which satisfies $x_\text{min} \le x_{s,k}^{(0)} \le x_\text{max}$ and $y_\text{min} \le y_{s,k}^{(0)} \le y_\text{max}$. 
	It is known that the target of the SA-PSO method is to find the best position among these particles and let it be a solution of \textbf{P3}. The solving process of \textbf{P3} based on the SA-PSO method is summarized in Algorithm \ref{AlgorithmA}. The introduction of Algorithm \ref{AlgorithmA} is given as follows.

	\subsubsection{Definition of the fitness function} To evaluate the fitness
of each particle, i.e., evaluating whether the positions of these particles can lead to satisfactory performance,
we first define the fitness function as 
	\begin{align}
	\label{FitFun}
\mathcal{F}(\bm{w}^*,\bm{p}_s^{(q)}) = \tilde{R}_p(\bm{w}^*) - r_1 \left| \mathcal{R}(\bm{p}_s^{(q)})\right|.
	\end{align}
 In \eqref{FitFun},  $\tilde{R}_p(\bm{w}^*)$ is the maximum primary  rate derived by solving \textbf{P2.2}  in Section \ref{OpTB} with the fixed positions of MAs, $\bm{p}_s^{(q)}$ represents the updated positions of the $q$-th iteration of implementing the SA-PSO method, $ \mathcal{R}(\bm{p}_s^{(q)})$ is a set including the pair positions of MAs violating the constraint C4 and defined as 
	\begin{align}
\left| \mathcal{R}(\bm{p}_s^{(q)})\right| = \{(\bm{p}_k, \bm{p}_\iota) | \left \| \bm{p}_k-\bm{p}_\iota \right\|_2 < d_\text{min}, 1\le k < \iota \le K \},
	\end{align}
	where $\left| \mathcal{R}(\bm{p}_s^{(q)})\right|$ represents the cardinality of $\mathcal{R}(\bm{p}_s^{(q)})$. Moreover, $r_1$ is the penalty factor and  set to be sufficiently large to satisfy  $\tilde{R}_p(\bm{w}^*) -r_1 \le 0$. This setting ensures that the constraint C4 can be finally met, i.e., $\left| \mathcal{R}(\bm{p}_s^{(q)})\right|$ will trend to be zero with the iterations. 

	\subsubsection{Updating of the particles' positions and  velocities} The updating of the $s$-th particle's position is controlled by the locally optimal position  of itself, i.e., $\bm{p}_{s,\text{best}}^{(q)}$, and the globally optimal position among all the particles, i.e., $\bm{p}_{\text{best}}^{(q)}$, which are evaluated by using the fitness function \eqref{FitFun}. According to \cite{PSOMA}, the velocity and position of the $s$-th particle are updated based on \eqref{UpdateV} and \eqref{UpdateP}, which are given by 
	\begin{align}
	\label{UpdateV}
\bm{v}_s^{(q+1)} &= \omega \bm{v}_s^{(q)} +c_1 r_2 (\bm{p}_{s,\text{best}}^{(q)} -\bm{p}_s^{(q)} ) + c_2 r_3 (\bm{p}_{\text{best}}^{(q)} -\bm{p}_s^{(q)} ), \nonumber  \\ 
& ~~~~~~~~~~~~~~~~~~~~~~~~~~~~~~~~~~~~~~~ s= 1,\ldots, S,\\
\label{UpdateP}
\bm{p}_s^{(q+1)} &= \bm{p}_s^{(q)} + \bm{v}_s^{(q+1)}, ~~~~~~~~~~~~~~~~~~s= 1,\ldots, S,
	\end{align}
	where  $\omega$ is the inertia weight parameter,  $c_1$ and $c_2$ are the step size factors, $r_2$ and $r_3$ are random parameters and uniformly generated from  $[0,1]$. If the updated positions violate the constraint C3, the following operation will be implemented as \cite{PSOMA}

 \begin{equation}
 	  	\label{CheckP}
 	  	z_{s,k}^{(q)}  = \left \{
 	  	\begin{aligned}
 	  	       &z_\text{min},  ~~~&\text{if}~~ z_{s,k}^{(q)} < z_\text{min},\\
             & z_\text{max}, ~~~ &\text{if}~~ z_{s,k}^{(q)} > z_\text{max},\\
              & z_{s,k}^{(q)},  ~&\text{otherwise},
 	  	\end{aligned}
 	  	\right.
 	  \end{equation}
 	  where $z_{s,k}^{(q)} \in \{ x_{s,k}^{(q)}, y_{s,k}^{(q)} \}$. The update of the particles' positions and velocities is presented in step 6 of Algorithm \ref{AlgorithmA}.

 	  \subsubsection{Updating of the globally optimal position}\label{SA}To avoid the solution space shrinking and the deterioration of solution accuracy suffered by the traditional PSO method, the SA method is adopted here, based on which a worse solution than the current one may be accepted with the probability $\epsilon$. This operation can prevent the solving process from being limited to the region of locally optimal solutions.  
 	  It is known that an appropriate setting
 	 of $\epsilon$ is  important for convergence performance and solution accuracy. According to \cite{HPSO}, $\epsilon$ can be updated as 
 	 \begin{align}
 	 \label{Probability}
            \epsilon = \exp \left(\frac{\mathcal{F} (\bm{w}^*,  \bm{p}_{\text{best}}^{(q)} )- \mathcal{F} (\bm{w}^*,  \bm{p}_{\text{best}}^{(q+1)} )  } {T^{(q)}} \right),
 	 \end{align}
where $T^{(q)}$ is the system  temperature  in the $q$-th iteration and updated as
$T^{(q+1)} = \frac{Q-q}{Q} T^{(q)}$, 
where $Q$ is the maximum number of iterations for the SA-PSO method.	It is known that the setting of $T^{(0)}$ is empirically.  This update of the globally optimal position is shown in steps 13-23 of Algorithm \ref{AlgorithmA}. Specifically, in step 13, we sort	$\bm{p}_s^{(q+1)}$ for $s =1,\ldots,S$ in the decreasing order in terms of its corresponding fitness function $\mathcal{F}(\bm{w}^*,\bm{p}_s^{(q+1)})$. From steps 16-20, the decision of whether accepting a worse solution than the current one is made.
$\bm{p}_{\varpi}^{(q+1)} $ defined in step 17 of Algorithm \ref{AlgorithmA} represents the $\varpi$-th particle in the $(q+1)$-th iteration, and $\varpi$ is randomly generated from the region  $[0, q+1]$.


  	 \begin{algorithm}
  	\caption{ The SA-PSO based method for solving \textbf{P3}}
  	\label{AlgorithmA}
  	\begin{algorithmic}[1]  
  		\STATE {Initialization:  Positions $\mathcal{P}^{(q)}$, velocities $\mathcal{V}^{(q)}$, the iteration index $q=0$, the maximum iteration number $Q$, the system temperature $T^{(q)}$.}
  		\STATE{Calculate $\mathcal{F} (\bm{w}^*,\bm{p}_s^{(q)})$ based on \eqref{FitFun}.} 
  		\STATE{Let  $\bm{p}_{s,\text{best}}^{(q)} = \bm{p}_s^{(q)}$ and $\bm{p}_{\text{best}}^{(q)} = \arg \max \{ \mathcal{F} (\bm{w}^*,\bm{p}_1^{(q)}), \ldots, \mathcal{F} (\bm{w}^*,\bm{p}_S^{(q)}) \}$.}
  		\WHILE{$q < Q$} 
  		\FOR{$s=1:S$}
  			\STATE {Update $\bm{v}_s^{(q+1)}$  based on \eqref{UpdateV}, and update $\bm{p}_s^{(q+1)}$ based on \eqref{UpdateP} and \eqref{CheckP}. } 
            \IF{$\mathcal{F} (\bm{w}^*,\bm{p}_s^{(q+1)}) > \mathcal{F} (\bm{w}^*,\bm{p}_{s,\text{best}}^{(q)}) $}
                \STATE{Let $\bm{p}_{s,\text{best}}^{(q+1)} = \bm{p}_s^{(q+1)}$.}
            \ELSE{}
                 \STATE{Let $\bm{p}_{s,\text{best}}^{(q+1)} = \bm{p}_{s,\text{best}}^{(q)}$.}
            \ENDIF
         \ENDFOR
                \STATE{Sort $\bm{p}_s^{(q+1)}$ for $s =1,\ldots,S$ in the decreasing order in terms of its corresponding fitness function $\mathcal{F} (\bm{w}^*,\bm{p}_s^{(q+1)})$, and let $\bm{p}_\text{temp} =\bm{p}_1^{(q+1)}$.}
                \IF{$\mathcal{F}(\bm{p}_\text{temp}) < \mathcal{F}(\bm{p}_\text{best}^{(q)})$  }
                    \STATE{Update $\epsilon$ based on \eqref{Probability}.}
                    	\IF{$\epsilon > \text{rand}(0,1)$} 
  		                      \STATE{Update $\bm{p}_\text{best}^{(q+1)}$ by letting $\bm{p}_\text{best}^{(q+1)} = \bm{p}_{\varpi}^{(q+1)}$. }
  		                 \ELSE{}
  		                      \STATE{$\bm{p}_\text{best}^{(q+1)} = \bm{p}_\text{best}^{(q)}$.}
  		                 \ENDIF
                    \ELSE{}
                    \STATE{$ \bm{p}_\text{best}^{(q+1)} = \bm{p}_\text{temp}$.  }
                \ENDIF
  		  \STATE{$q=q+1$.}
  		\ENDWHILE
  	    \STATE{Return $\bm{p}$.}
  	\end{algorithmic}
  \end{algorithm}

\subsection{The proposed AO algorithm for solving \textbf{P1}}
In this subsection, we summarize the solving process of \textbf{P1} in Algorithm \ref{AlgorithmB} based on the descriptions in Section \ref{OpTB} and \ref{OpMA}. The computational complexity of Algorithm \ref{AlgorithmB} is $\mathcal{O}(A_1 SQ+ A_1 A_2K^{4.5} \log(\frac{1}{\hat{\epsilon}}      ))$, where $A_1$ is the iteration number of implementing the AO framework, $A_2$ is the iteration number of implementing the SCA method for \textbf{P2.2}, and $\hat{\epsilon}$ is the computational accuracy required for implementing the interior-point method. It is noted that the proposed algorithm with the PSO can also be used to solve \textbf{P1}, the complexity of which is approximately as $\mathcal{O}(A_1 SQ+ A_1 A_2K^{4.5} \log(\frac{1}{\hat{\epsilon}}      ))$. The performance comparison between the proposed scheme with the two algorithms will be conducted in Section \ref{section-IV}.

  	 \begin{algorithm}
  	\caption{ The AO algorithm for solving \textbf{P1}}
  	\label{AlgorithmB}
  	\begin{algorithmic}[1]  
  		\STATE {Initialization:  The  MAs' positions $\bm{p}^{(0)}$, the AO iteration index $\xi=0$,  the convergence tolerance $\hat{\varrho}$.}
  		\REPEAT
           \STATE{$\xi=\xi+1$.}
           \STATE{Initialization: the SCA iteration index $\varsigma=0$, the transmit beamforming matrix $\bm{W}^{(\varsigma)}$.}
           \REPEAT
                \STATE{$\varsigma=\varsigma+1$.}
                \STATE{Update $\bm{W}^{(\varsigma)}$ by solving \textbf{P2.2}.}
           \UNTIL{$|\tilde{R}_p^{(\varsigma)} - \tilde{R}_p^{(\varsigma-1)} | \le \hat{\varrho}$.}
           \STATE{Obtain $\bm{w}^{(\xi)}$ by implementing the singular value decomposition of $\bm{W}^{(\varsigma)}$.}

           \STATE{Update $\bm{p}^{(\xi)}$ by implementing Algorithm \ref{AlgorithmA}.}
  		\UNTIL{ $|R_p^{(\xi)} -R_p^{(\xi-1)} | \le \hat{\varrho}$. }
       \STATE{Return $\bm{w}$ and $\bm{p}$.}
  	\end{algorithmic}
  \end{algorithm}

\section{Numerical Results}\label{section-IV}
In this section, numerical results are conducted to evaluate
the performance of the proposed MA empowered scheme with the SA-PSO. We consider a two-dimensional topology, in which the locations of  the BD and the PU are set at  (30 m, 40 m) and ($\hat{\xi}$, 0), respectively, and $\hat{\xi}$ is uniformly generated between 30 m and 60 m. The moving region for MAs at the PT is considered as a square area of size $[-\frac{A}{2},\frac{A}{2}]\times [-\frac{A}{2},\frac{A}{2}]$ at the center of (0 m, 0 m), where $A = 3\lambda$, and $\lambda$ is the carrier wavelength  and set at 0.1 m.  The amount of transmit/receives paths is set to be the same, i.e., $L_p^t = L_s^t = L_\kappa^r= L$ \cite{MATWC}.  The path-response matrices $\bm{\Sigma}_\kappa$ and $\bm{\Sigma}_{s}$ can be expressed as $\text{diag}\{[\widehat{n}_{\kappa,1}, \ldots, \widehat{n}_{\kappa,L}]\}$ and $\text{diag}\{[\widehat{n}_{s,1}, \ldots, \widehat{n}_{s,L}]\}$, respectively, where $\widehat{n}_{\kappa,\overline{l}}$ and $\widehat{n}_{s,\overline{l}}$  are the corresponding complex responses of the $\overline{l}$-th path for $\overline{l}= 1, \ldots, L$ and  satisfy $\mathcal{CN}(0,\upsilon\cdot d^{-\nu}/L)$. $\upsilon$ denotes the path-loss, $d$ represents the distance between two nodes, and  $\nu$ is the pass-loss exponent.  The AoAs and AoDs are  uniformly generated from $[-\frac{\pi}{2},\frac{\pi}{2}]$. The remaining parameters are set as follows: $P_\text{max} = 38$ dBm, $\alpha =0.8$, $d_\text{min} = 0.5\lambda$, $\upsilon=-10$ dB, $\nu = 1.8$, $L=9$, $Q = 150$, $S = 150$, $c_1 = c_2 = 1.4$, $\omega = 1.2$, $r_1 = 50$, $\sigma^2= 10^{-8}$, and $\hat{\varrho} = 10^{-2}$. For performance comparison, the proposed MA empowered scheme with the PSO, the FA empowered scheme and the random transmit beamforming scheme  are exploited.

\begin{figure}[t]
  \centering
  \includegraphics[width=0.35\textwidth]{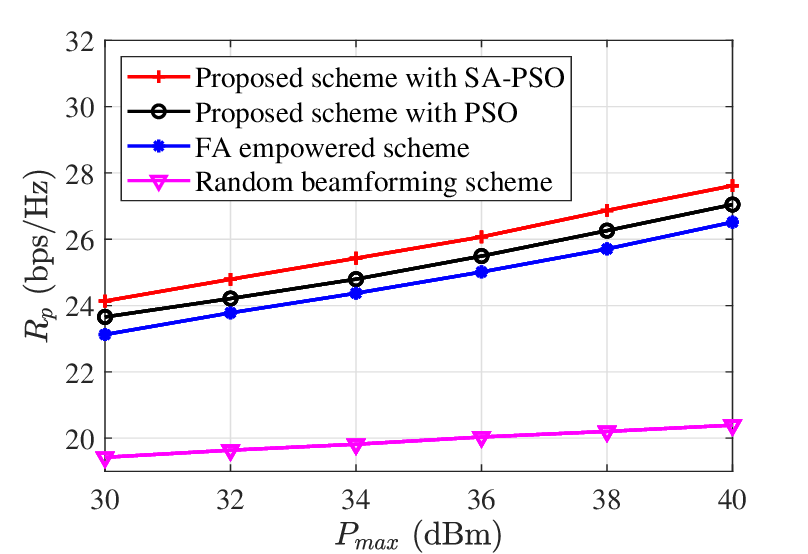}
  \caption{Primary  rate versus maximum transmit power.}
  \label{fig:power}
\end{figure}

Fig. \ref{fig:power} investigates the variance of
the primary  rate versus the maximum transmit power. It is obvious that increasing the maximum transmit power results in the improvement of the primary rates for all schemes. As indicated in Fig. \ref{fig:power}, the proposed scheme with the SA-PSO  can achieve a higher primary rate than the proposed scheme with the PSO, which confirms that the modification on the PSO method by utilizing the SA can guarantee a satisfactory solution space and thus find a solution with better accuracy. Compared to the FA empowered scheme, the utilization of MAs at the PT can ensure better system performance. It is because moving the antenna positions can  create  favorable channel conditions between the PT and the PU and between the PT and the BD.

\begin{figure}[t]
  \centering
  \includegraphics[width=0.35\textwidth]{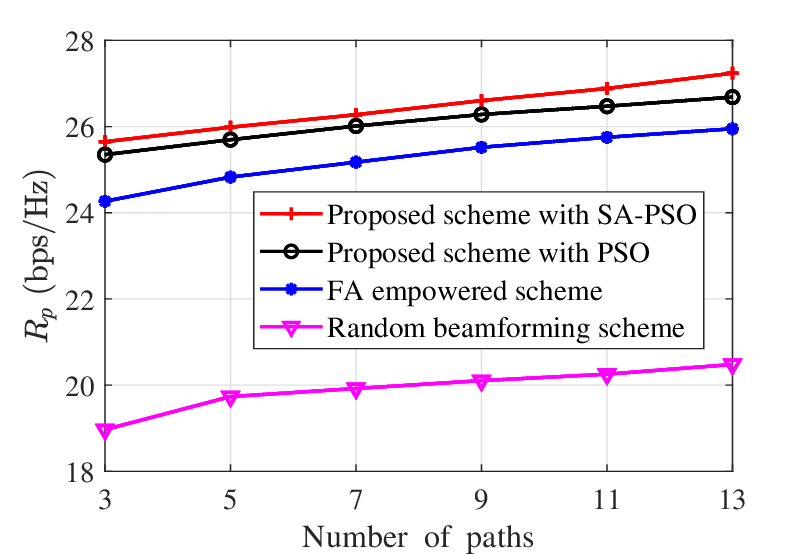}
  \caption{Primary  rate versus number of paths.}
  \label{fig:paths}
\end{figure}

Fig. \ref{fig:paths} shows how the number of transmit/receive paths affects the primary  rate. It is seen that the existence of more paths has a positive effect on the improvement of the primary  rate. This is due to the fact that more paths can ensure a greater multi-path gain, thereby enhancing the small-scale fading. From Fig. \ref{fig:paths}, we also find that optimizing the transmit beamforming  is a significant factor affecting the primary  rate. For example, compared to the FA empowered scheme, the random beamforming scheme even  results in a 21.8\% performance loss with $L=3$. It is because without the beamforming design, the transmitted signal from the PT will not beam towards the PU and the BD, thereby reducing the received signal power at the PU.

\begin{figure}[t]
  \centering
  \includegraphics[width=0.35\textwidth]{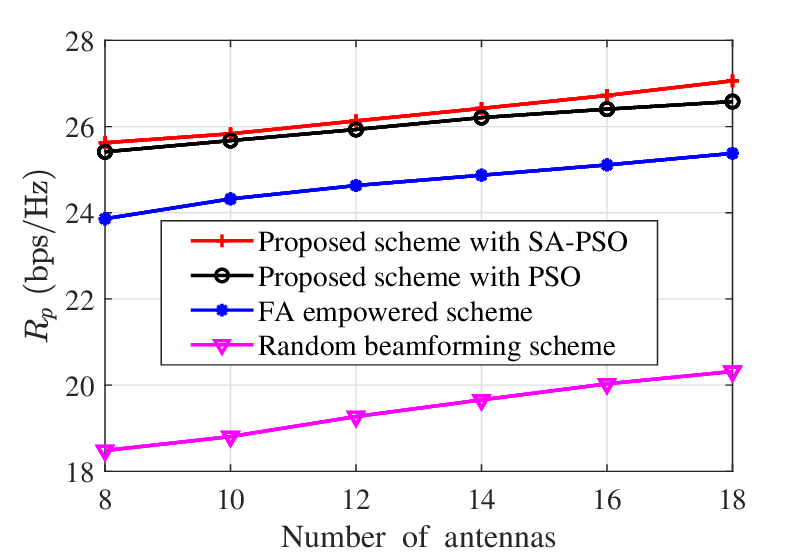}
  \caption{Primary  rate versus number of antennas.}
  \label{fig:antennas}
\end{figure} 

\begin{figure}[t]
  \centering
  \includegraphics[width=0.35\textwidth]{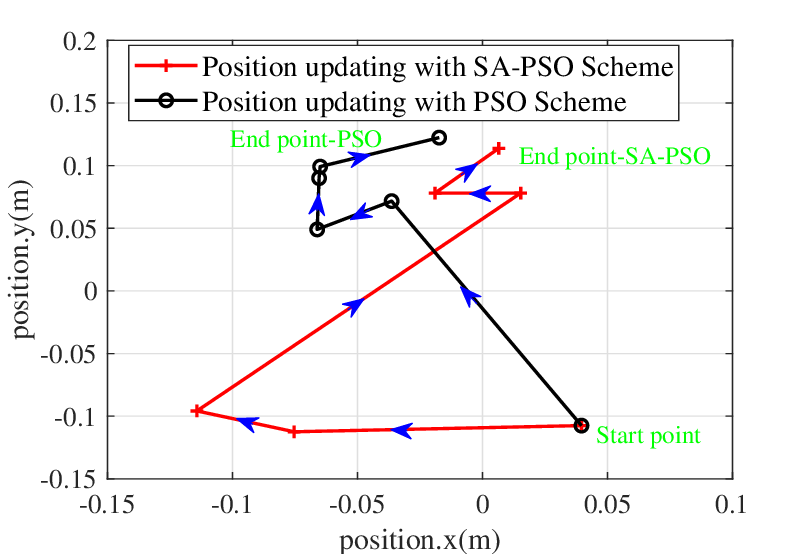}
  \caption{Position updating process of a selected antenna with AO iterations.}
  \label{fig:positions}
\end{figure}

Fig. \ref{fig:antennas}  depicts the primary  rate versus the number of antennas. By deploying more antennas at the PT, the primary  rates of all schemes increase since an improved spatial diversity gain can be achieved. Similar to Figs. \ref{fig:power} and \ref{fig:paths}, our proposed scheme with the SA-PSO outperforms the benchmarks, which again demonstrates the effectiveness of utilizing the MA technology and the SA-PSO method to ensure favorable channel conditions.

Fig. \ref{fig:positions} presents the position variation of a selected antenna versus the AO iterations. As shown in Fig. \ref{fig:positions}, the updating paths of the selected antenna for the SA-PSO and PSO are much different. Obviously, the SA-PSO  method can provide a larger solution space for updating the antenna position. Moreover, we can find that the AO framework can converge after only  five iterations. This observation indicates  the superior convergence performance of the proposed  algorithm. 

\section{Conclusion}
In this paper, we have proposed an MA empowered scheme to create favorable channel conditions to boost the transmission efficiency of SR communication systems. To optimize the primary transmission rate under the BER constraint on the secondary transmission, we have proposed an AO framework to iteratively optimize the transmit beamforming by using the SCA and SDP methods and optimize the positions of MAs by using the SA-PSO method. Finally, we have conducted numerical results to demonstrate the effectiveness of the proposed MA empowered scheme with the SA-PSO. From numerical results, we have derived the the following observations: 1) the flexible positions of antennas at the PT can provide a higher spatical diversity for both primary and secondary transmissions; 2) creating a multi-path environment has a positive effect on system performance.

\end{document}